\documentclass[aps,pra,twocolumn,showpacs,superscriptaddress]{revtex4}  
\usepackage{graphicx}  % needed for figures
\usepackage{dcolumn}   % needed for some tables
\usepackage{bm}        % for math
\usepackage{amssymb}   % for math
\usepackage{textcomp}

\begin{document}

\title{Quantum diffusion and thermalization at resonant tunneling}

\author{Carlos~A.~Parra-Murillo}
\affiliation{Institut f\"ur Theoretische Physik, Universit\"at Heidelberg, 69120 Heidelberg, Germany}
\author{Javier~Madro\~nero}
\affiliation{Departamento de F\'isica, Universidad del Valle, Cali, Colombia}
\author{Sandro Wimberger}
\affiliation{Institut f\"ur Theoretische Physik, Universit\"at Heidelberg, 69120 Heidelberg, Germany}
\affiliation{Dipartimento di Fisica e Scienze della Terra, Universit\`a di Parma,  Via G. P. Usberti 7/a, 43124 Parma, Italy}
\affiliation{INFN, Sezione di Milano Bicocca, Gruppo Collegato di Parma, Italy}

%\date{\today}

\begin{abstract}
Nonequilibrium dynamics and effective thermalization are studied in a resonant tunneling scenario via multilevel Landau-Zener crossings. Our realistic many-body system, composed of two energy bands, naturally allows a separation of degrees of freedom. This gives access to an effective temperature and single- and two-body observables to characterize the delocalization of eigenstates and the non-equilibrium dynamics of our paradigmatic complex quantum system.
\end{abstract}


\maketitle

\section{Introduction}

Complex systems and their modeling are interesting simply because matter is typically complex. On small
scales, quantum mechanics  describes interactions and dynamics. One typical example of a complex 
quantum system are photosynthetic molecules (complexes) in which transport properties seem to be governed by quantum mechanical interference \cite{BuchleitScholes2012}. Strong correlations between the constituents make even "small" systems very complicated. In the Helium atom, for instance, classical three-body chaos turns into complicated ionization spectra \cite{TannerRichterRMP2000}. 

The modern experimental tools of atom optics allow for a bottom-up construction of strongly correlated many-body quantum system \cite{JakschZoller2005}.
Here we study a lattice model for bosons hitherto hardly investigated \cite{Morebands,Exptwobands}. 
Our two-band Bose-Hubbard model shows all the ingredients of a complex quantum system, is realistically
implemented with ultracold atoms, see \cite{CarlosPRA2013}, and acts as versatile toolbox to study and control many-body quantum 
evolutions \cite{Ploetz2010,CarlosPRA2013}. After introducing the model and a brief review of its spectral properties, we show how thermalization
and many-body localization \cite{StatMecFound,ThermExp1,FazzioSantos} in this isolated
 quantum system depends on its (non)integrability. The latter can be controlled
by tunable system parameters; here we use the interparticle interaction strength and a tilt force coupling the  
two bands. The two bands allow for a natural division into subsystems whose entanglement properties are studied. The purity of 
the quantum state, which we use to do so, relates to an effective temperature of the subsystems. 
Finally, we sweep the tilt to investigate the evolution of one- and two-body observables and their thermalization in the course of time.

\section{Two-band Bose-Hubbard system}

Our analysis of the spectral (static) and dynamical features is based on the two-band Bose-Hubbard Hamiltonian: 
\begin{eqnarray}\label{eq:01}
 \hat{H} &=&\sum_{l,\beta}\left[-\frac{J_{\beta}}{2}\left(\hat{\beta}^{\dagger}_{l+1}\hat{\beta}_{l}\,e^{-i2\pi Ft}+h.c.\right)+
 \frac{gW_{\beta}}{2}\hat{\beta}^{\dagger 2}_{l}\hat{\beta}^2_{l}\right]\,\nonumber\\
 &+&\sum_{l,\mu}2\pi F C_{\mu} \hat{a}^{\dagger}_{l+\mu}\hat{b}_l\,e^{-i2\mu\pi Ft}+
\frac{gW_x}{2}\hat{b}^{\dagger 2}_{l}\hat{a}^2_{l}+h.c.\nonumber\\
 &+&\sum_l2gW_x\hat{n}^a_l\hat{n}^b_l+\frac{\Delta}{2}(\hat{n}^{b}_l-\hat{n}^{a}_l)\,,
\end{eqnarray}
\noindent where $\hat \beta_l(\hat \beta^{\dagger}_l)$ represents the annihilation (creation) operators, and $n^{\beta}_l=\hat\beta^{\dagger}_l\hat\beta_l$
is the number operator, with band index defined as $\beta=\{a,b\}$. $J_{\beta=a,b}$ are the hopping matrix elements, $gW_{a,b,x}$ intra- and interband
particle-particle interaction strengths, with $g\in[0,1]$. $C_{\mu}$ represents interband coupling terms induced by the force or, equivalently, by the lattice acceleration $F$ \cite{WimbPRL2007,MGreiner2011}. $\Delta$ is the band gap, which is controlled by the actual physical implementation \cite{CarlosPRA2013}.

Here we study the behavior of our system in the parameter space $(g,F)$.
We impose periodic boundary conditions in space, i.e. $\hat{\beta}^{\dagger}_{L+1}=\hat{\beta}^{\dagger}_{1}$. 
This allows us to work with the translationally invariant Fock basis (TIFB) defined as in \cite{kol68PRE2003}: 
 $|s_{\alpha},\kappa\rangle=D_{\alpha}^{-1/2}\sum^{D_{\alpha}}_{k=1}\hat
 e^{-i2\pi \kappa l}\hat S^k|\vec{n}_a\rangle_{\alpha}\otimes|\vec{n}_b\rangle_{\alpha}$,
 where $\hat S=\hat S_a\otimes\hat S_b$ is the translation operator of the composite Hilbert space $\mathcal H_s=\mathcal H_b\otimes\mathcal H_b$.
The respective dimension is given by $\mathcal{N}_s=(N+2L-1)!/LN!(2L-1)!$, for $N$ atoms in $L$ sites. Without loss of generality we work here with the subspace defined by the quasimomentum $\kappa=0$, resulting in a reduction of the Fock space dimension by a factor of $L$ \cite{CarlosPRA2013,kol68PRE2003}. 

%******** FIG 1
\begin{figure}[t]
\centering \includegraphics[width=0.95\columnwidth]{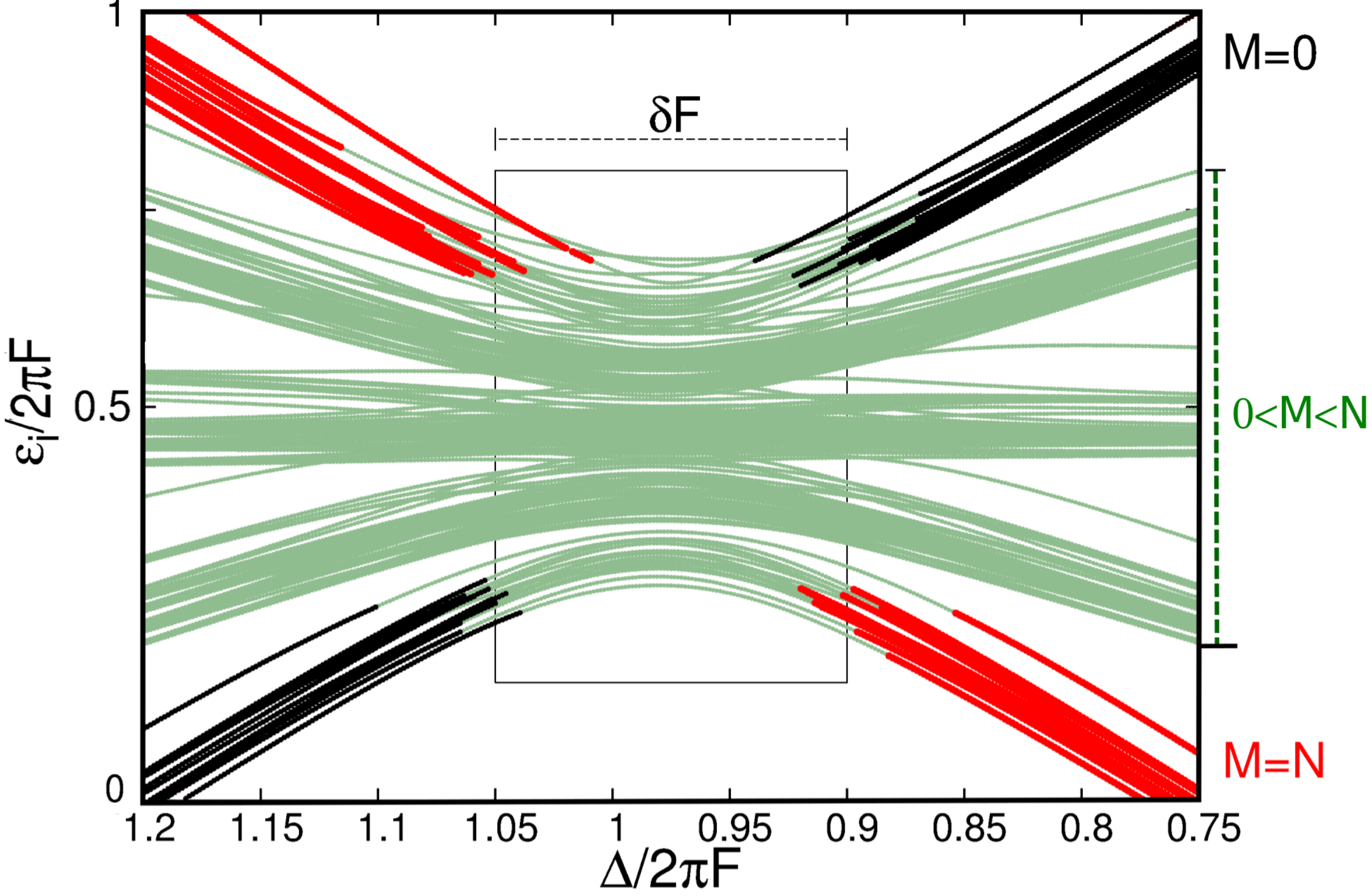}
\caption{\label{fig:1}(Color online): Many-body spectrum around $F_{r=1}$ and $N/L=4/5$. 
The different color lines correspond to eigenstates of the type: $|\psi_{M=0}\rangle$ (black  lines), $|\psi_{M=N}\rangle$ (red thick
lines), and the mixed-like states, i.e., states with $0<M<N$ are represented by the green thin lines. $\delta F$ represents the 
strongly coupled spectral region characterized by avoided crossings. The parameters are chosen for best visibility such that the
spectrum is nearly regular. They follow from an implementation with a double-periodic lattice \cite{CarlosPRA2013}: $\Delta=1.61$, $J_a=0.082$, $J_b=-0.13$, $C_0=-0.094$, $C_{\pm 1}=0.037$, $C_{\pm  2}=-0.0022$. $W_a=0.021$, $W_b=0.026$ and $W_x=0.023$.}
\end{figure}
%******** FIG 1

Our system shows a transition from regular to complex (chaotic) quantum spectra in the vicinity of tunneling resonances \cite{CarlosPRA2013}. They occur if the tilt compensates exactly the band gap. Resonantly enhanced tunneling (RET) was studied experimentally in the mean-field regime of our model \cite{WimbPRL2007} as well as in single-band \cite{MGreiner2011} many-body setups. At RET, i.e. at specific values $F=F_r$ \cite{Ploetz2010}, the interband coupling is maximized. $r$ is the order of the resonance, and for simplicity we fix $r=1$ here. Because of our acceleration gauge, $\hat H$ is periodically time-dependent. We naturally use the Floquet approach \cite{Shirley1965} to study the quasi-energies of the system. All energies are measured in recoil energies \cite{JakschZoller2005} and the lattice constant is $2\pi$. Then the quasi-energies lie in the Floquet zone (FZ), $\varepsilon_i\in[0,2\pi F]$, and the extended spectra are obtained by $\varepsilon_{i,n}=\varepsilon_i+2\pi n F$, with $n\in \mathbb{Z}$. The numerical implementation is challenging because of the size of the involved Floquet matrices \cite{CarlosPRA2013}.

Fig.~\ref{fig:1} shows such a spectrum in the (quasi)integrable regime such that the spectral structure can be best
appreciated. The regime close to RET is market by $\delta F$. In the off-resonant regime, $F\notin [F_r-\delta F/2,F_r+\delta F/2]$, the
spectrum can be split into classes of eigenstates, labeled by the
upper-band  occupation number $M_i=\langle \sum_l\hat
n^b_l\rangle_{\varepsilon_i}$. The corresponding subsets are referred to as \emph{$M$ manifolds}. While these manifolds substantially
increase the complexity of the system with respect to the single-particle or mean-field Landau-Zener tunneling \cite{WimbPRL2007}, 
they offer great possibilities to study many-body quantum diffusion \cite{Wilkinson1988}. Our two-band model gives
a natural separation also into subbands. Their properties and the interband coupling can be investigated by the following two-body
correlation functions: 
\begin{equation}
\theta_{\beta}=\left\langle\frac{1}{2}\sum\nolimits_l\hat{\beta}^{\dagger  2}_{l}\hat{\beta}^2_{l}\right\rangle_{\varepsilon_i},
\;\;\text{and}\;\; \theta_x=\left\langle2\sum\nolimits_l\hat n^a_l\hat n^b_l\right\rangle_{\varepsilon_i}\,,
\end{equation}
with $\beta=\{a,b\}$. As shown in \cite{CarlosACTAPOL2013}, the spectrum is well described in the off-resonant regime by the quantum numbers  
$\{M,\theta_a,\theta_b,\theta_x\}$. This implies a nearly integrable system in this regime, characterized by regular spectral
correlations \cite{CarlosACTAPOL2013,CarlosPRA2013}. 

The non-integrability arises from the increasing degree of manifold mixing at RET. 
Therein, the loss of good quantum numbers is due to level repulsion, which induce chaotic spectral statistics. In fact, the
interplay between interaction ($g$) and resonant tunneling ($F_r$) induces a transition from a regular to quantum chaotic
spectrum. The conditions for quantum chaos at RET are: $N/L\sim1$ and $2\pi F r\approx\Delta\leq1$, and $g>0.5$ (see ref.~\cite{CarlosPRA2013}).

\section{Spectral diffusion}

We come now to our main purpose, the study of spectral diffusion and non-equilibrium dynamics. 
Our system is perfect for this scope, since we can drive an initial state $|\psi_0\rangle$ across the  RET regime where non-adiabatic 
transitions take place. For this, we use a linear sweeping function $F(t)=\dot{F}t+F_0$, $t\geq 0$.

\begin{figure}[t]
\centering \includegraphics[width=\columnwidth]{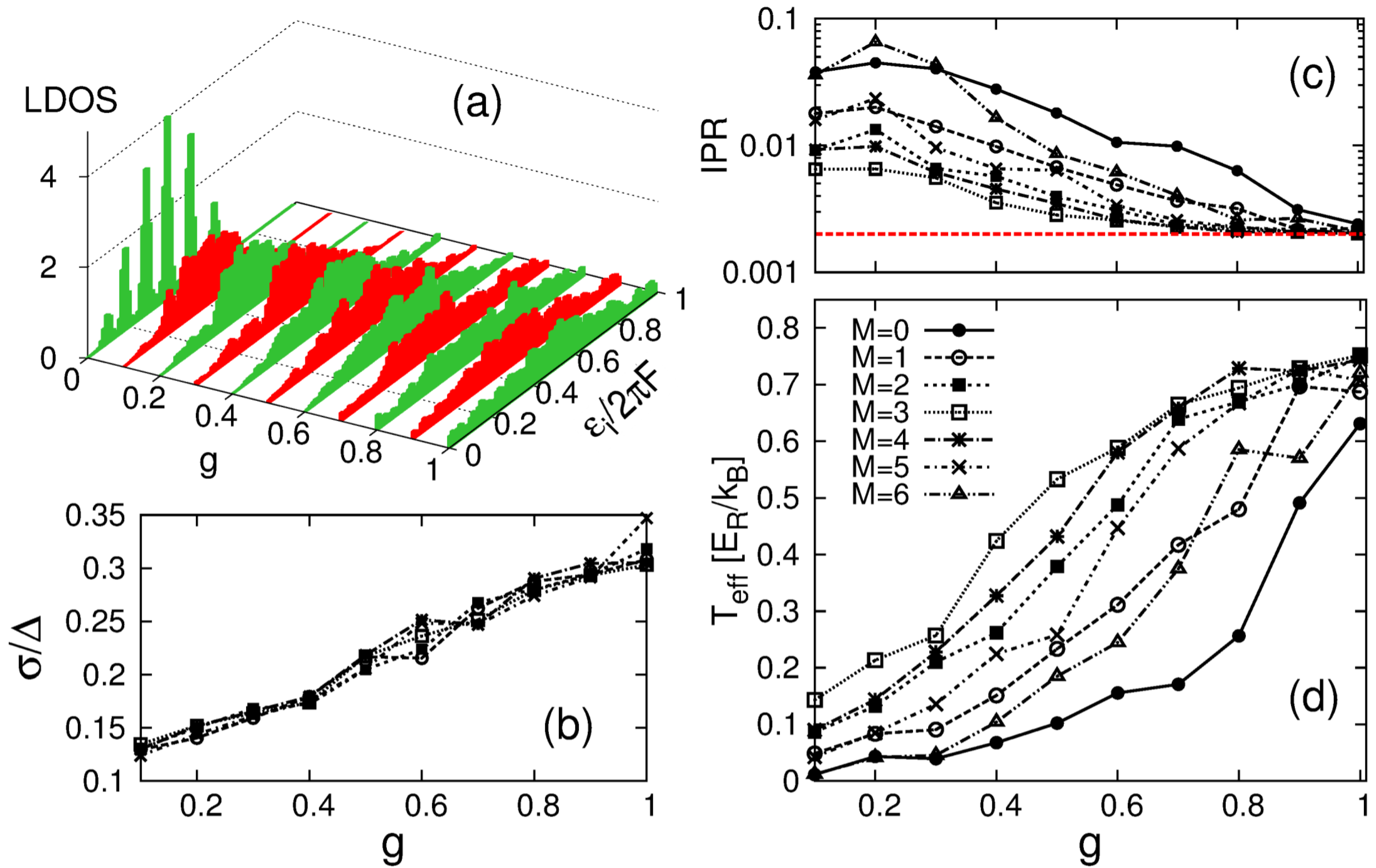}
\caption{\label{fig:2}(Color online): (a) Diffusive spreading of the LDOS at $F_{r=1}$ ($N/L=6/5$) vs 
 $g$. The initial state is: $|\varepsilon_i\rangle\approx \sum_k\hat S^k|01110\rangle\otimes|01110\rangle$ at $F=\Delta/3\pi$. 
(b) standard deviation $\sigma$ within
the manifold $0< M < N$. (c) IPR in TIFB. The dashed-line is the statistical limit of
ref.~\cite{CarlosPRA2013} (there called random matrix theory (RMT) limit). (d) Effective temperature computed by means 
of the purity. $M$ denotes the manifold number defined in text. 
The parameters are: $\Delta=0.28$, $J_a=0.038$, $J_b=-0.042$, 
$C_0=-0.097$, $C_{\pm 1}=0.046$, $C_{\pm 2}=-0.0008$, $W_a=0.028$, $W_b=0.029$, and $W_x=0.0286$, for which the spectrum is chaotic around $F_r$.} 
\end{figure}

The presence of avoided crossings (ACs) in the spectrum, see Fig.~\ref{fig:1}, generates a spreading of the wave packet in the
instantaneous basis of Floquet states with energies $\varepsilon_i(F(t))$. For increasing interaction strength $g$, more
and more ACs appear until the spectrum becomes fully chaotic \cite{CarlosPRA2013}.
The local density of states (LDOS), defined by $P_{\psi}(\varepsilon,g)=\sum_i|C_i|^2\delta(\varepsilon-\varepsilon_i)$,  
with $C_i\equiv \langle \psi_t|\varepsilon_i\rangle$, characterizes 
this spreading (see Fig.~\ref{fig:2}(a)). The variance of this probability distribution,
see Fig.~\ref{fig:2}(b), grows almost linearly with $g$, until a nearly flat distribution is reached within the FZ. Here $|C_i|^2\sim
1/\mathcal{N}_s$, i.e. the system obeys a equipartition condition. This can be seen also from the Shannon information entropy 
\cite{Smilansky1987}: $S_{\rm sh}=-\sum_i|C_i|^2\ln |C_i|^2$,  approaching $S_{\rm  sh}\approx\ln \mathcal{N}_s$ in
statistical equilibrium. We come back to $S_{\rm sh}$ at the end of the paper when studying the reversibility of this equilibration process. 

% ******** FIG 2
%***************

An alternative way to describe the interband mixing is offered by analyzing the subsystems of the total Hilbert space
$\mathcal{H}_s=\mathcal{H}_a\otimes\mathcal{H}_b$ provided by the two bands. To do so, we look at the reduced density operator
associated with either of the bands after tracing out the other one. The trace is best performed with the help of the following single
band states, shifted by $k$ positions in Fock space, $|a(b)_{k,\alpha}\rangle=\hat S^k_{a(b)}|\vec n_{a(b)}\rangle_{\alpha}$, 
since in general $|s_{\alpha}\rangle\neq|s^a_{\alpha}\rangle\otimes|s^b_{\alpha}\rangle$, with $|s^{\beta=a,b}_{\alpha}\rangle$ being a single-band TIFB state. 
The density operator of the evolving state,
$\hat{\rho}_t=|\psi_t\rangle\langle\psi_t|=\sum_{\alpha,\beta}A_{\alpha}(t)A^*_{\beta}(t)|s_{\alpha}\rangle\langle s_{\beta}|$, 
can then be written in this basis as:
$\hat{\rho}_t=\sum_{kk',\alpha\beta}\Lambda^{\alpha\beta}_t|a_{k,\alpha}\rangle\langle a_{k',\beta}|\otimes|b_{k,\alpha}\rangle\langle b_{k',\beta}|$,
with $\Lambda^{\alpha\beta}_t=A_{\alpha}(t)A^*_{\beta}(t)(D_{\alpha}D_{\beta})^{-1/2}$. We now trace out the degrees of freedom
 $\mathcal H_b$, which results in
\begin{eqnarray}\label{eq:02}
\hat{\rho}^a_t=\sum\nolimits_{k,\alpha}\Lambda^{\alpha\alpha}_t|a_{k,\alpha}\rangle\langle a_{k,\alpha}|=
\sum\nolimits_{\alpha}\Lambda^{\alpha\alpha}_t\hat\rho_{\alpha\alpha}
\end{eqnarray}
where we have used $\sum_{p\lambda}\langle b_{p\lambda}|b_{k,\alpha}\rangle\langle b_{k',\beta}|b_{p\lambda}\rangle=\delta_{kk'}\delta_{\alpha\beta}$.
The reduced density operator is thus decomposed into a mixture of many-body states
$\hat\rho_{\alpha\alpha}=\sum_k|a_{k,\alpha}\rangle\langle a_{k,\alpha}|$, with fixed number of particles $0\leq N'\leq N$; it is 
straightforwardly proven that  ${\rm tr}(\hat{\rho}^a_t)=1$.  The mixedness of $\rho^{a}_t$ is measured by the purity
$\gamma[\hat \rho^{a}_t]\equiv {\rm  tr}\left((\rho^{a}_t)^2\right)$, which reads 
\begin{equation}\label{eq:03}
\gamma[\hat \rho^{a}_t]=\gamma[\hat \rho^{b}_t]=\sum\nolimits_{\alpha}|A_{\alpha}(t)|^4\,.
\end{equation}  
The result is nothing but the inverse participation ratio, ${\rm  IPR}=\sum_{\alpha}|A_{\alpha}(t)|^4$, 
in the TIFB, a well-known localization measure \cite{Dittich1991}. Therefore, a well-localized state in Fock space
has a large purity with an upper bound given by $\gamma=1$, whenever only one state of TIFB is populated. A fully mixed state 
has a minimal IPR and its purity is given by the statistical limit $\gamma\approx 2/\mathcal{N}_s$, 
where the equipartition condition $|A_{\alpha}|^2\sim 1/\mathcal{N}_s$ is fulfilled \cite{CarlosPRA2013}. 

Tracing over one energy-band, we can characertize the mixedness of the reduced state $\hat\rho^{a}_t$ by an effective temperature $T_{\rm eff}$ for the remaining degrees of freedom. For this, we use a sweep with $\dot{F}=d\delta F$, where $d$ defines the mean level spacing at $F_{1}$, we plot the IPR after
equilibration, at a time $t \geq t_1= (F_1-F_0)/\dot F$. An effective temperature is then defined by equating the numerically obtained
$\gamma[\hat \rho^{a}_t]={\rm IPR}$ with $\gamma[\hat\rho_{\rm Th}(\beta_{\rm eff})]$, where 
$\hat \rho^{a}_t\approx Z^{-1}\exp(-\beta_{\rm eff} \hat H'_a)$.  Here the normalization factor is given by the partition function 
$Z=Z_{\omega}\sum\nolimits_{i}\exp(-\beta_{\rm  eff}\varepsilon'_{i})$, with $Z^{-1}_{\omega}=1-\exp(-2\pi F\beta_{\rm eff})$, 
taking into account that we are dealing with a Floquet spectrum \cite{RKetzmerick2010}. $\hat H'_a$ is the Hamiltonian for
the band $a$ with a number of particles $0\leq N_a\leq N$. $\gamma[\hat\rho_{\rm Th}(\beta_{\rm eff})]=\gamma[\hat
\rho^{a}_t]= {\rm IPR}$ defines a non-linear equation for $\beta^{-1}_{\rm  eff}=k_BT_{\rm  eff}$, which is solved by a root
finding algorithm. 

In Fig.~\ref{fig:2}(c-d) we show the IPR and the effective temperature as a function of the interaction strength $g$. Like in Fig.~\ref{fig:2}(b), the results are averaged over 30 initial states within each manifold. The number of averaged initial states reduces fluctuations, otherwise it does not impact the outcome. The more states participate in the evolution, the larger is $\approx$ IPR$^{-1}$ and hence also $T_{\rm eff}$. $T_{\rm eff}$ essentially depends only on the manifold number of the initial conditions. For $\langle\psi_0|\hat M|\psi_0\rangle\approx N/2$, the spreading is faster since the coupling to the neighboring manifolds is more symmetric (see also Fig.~\ref{fig:1}). Therefore, the respective purity drops faster to $\gamma\approx 2/\mathcal{N}_s$ in this case than for initial states with $\langle\psi_0|\hat M|\psi_0\rangle\neq N/2$. The latter implies that the temperature is the higher the closer one starts to the center of the spectrum at $F_0$. Our proposal to introduce the effective temperature by the number of effectively coupled states overcomes the problem that a straightforward definition (as usually done in statistical mechanics, e.g. via the entropy) is highly non-trivial in driven systems because of strong fluctuations of the time-dependent quantities, see also the upcoming figures.

%**** FIG 3
\begin{figure}[t]
\centering \includegraphics[width=\columnwidth]{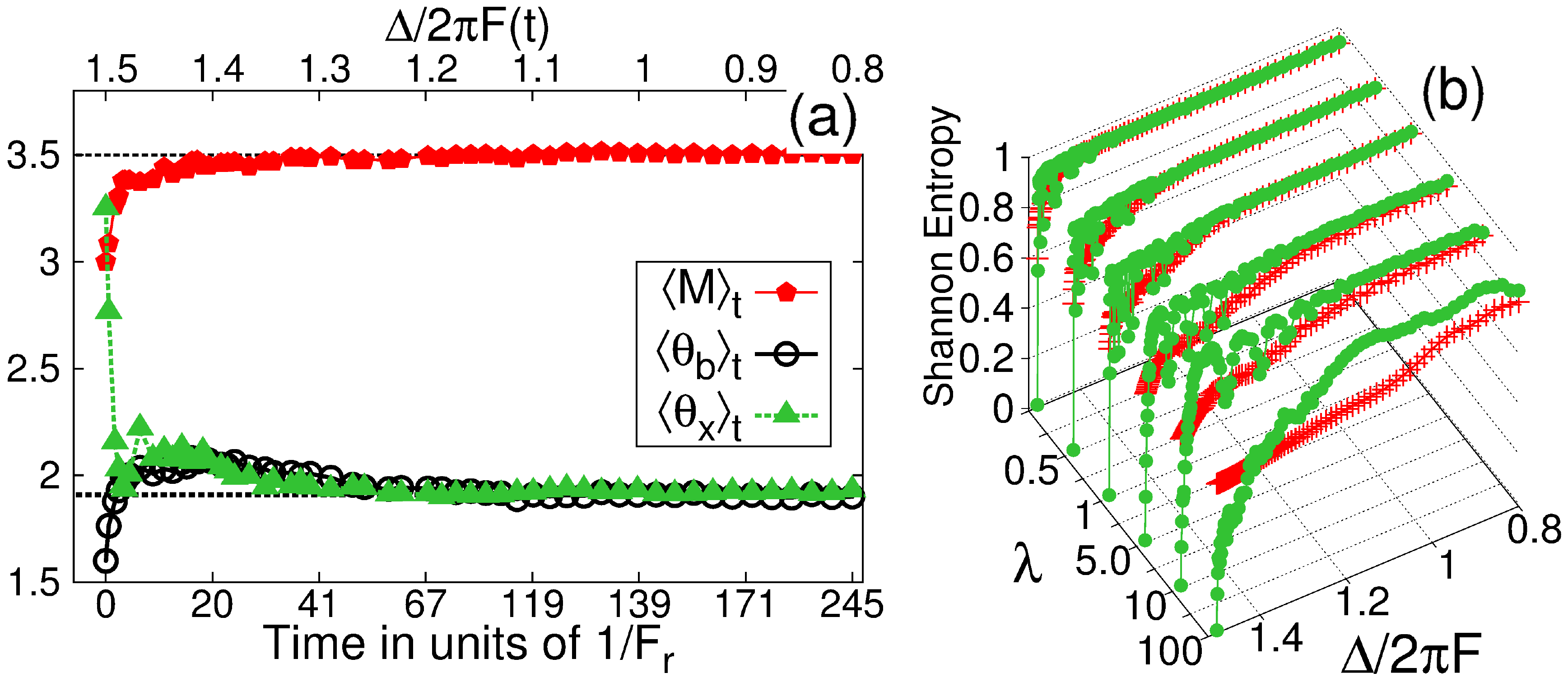}
\caption{\label{fig:3}(Color online): (a) Relaxation and thermalization of single- and two-particle 
observables for $g=1$ and $\lambda=0.25$. In panel (a-b) the evolution is defined by the pulse 
$F(t)=\dot F t+F_0$, and $\langle\psi_0|\hat M|\psi_0\rangle=3$. (b) Shannon entropy in both 
instantaneous (greed filled circles) and TIFB basis (red $\times$) vs ($\lambda$, $\Delta/2\pi F(t)$). 
The black dashed lines in (a) are the microcanonical averages at $F_{r=1}$ for $N/L=7/5$.
The parameters are the same as in Fig.~\ref{fig:2}.}
\end{figure}
%***************

The saturation value is obtained from the maximally mixed state reached at $g\approx 1$. For our many-body system,
finite-size effects have to be considered. As for the localization measures above, where
the dimension $\mathcal N_s$ defines a natural lower bound, the effective temperature will saturate to an upper bound depending on the
size of the accessible Hilbert space. This explains the behavior of the curves in Fig.~\ref{fig:2}(d) for $g\rightarrow 1$.

\section{Effective thermalization and irreversibility of quantum dynamics}

The thermalization of observables in a complex quantum system can be investigated on the basis of the eigenstate thermalization hypothesis \cite{StatMecFound}. First one checks that the expectation value of the corresponding operator $\hat O$ approaches its diagonal approximation in a finite evolution time, i.e. 
\begin{equation}\label{eq:04}
\langle\hat O\rangle_t \equiv \langle\psi_t|\hat O|\psi_t\rangle={\rm tr}(\hat  O\hat{\rho}_t) 
%&\sum_i|C_i|^2O_{ii}+\sum_{i\neq  j}|C_i||C_j|O_{ij}\exp(-i\Delta \phi_{ij})\nonumber\\
\rightarrow \sum_i|C_i|^2O_{ii}\,,
\end{equation}
with $O_{ij}=\langle \varepsilon_i|\hat O|\varepsilon_j\rangle$.  Secondly, we test whether the temporal average of an operator characterizing the system is approximately given by
\begin{equation}
\overline{O}=\lim_{t\rightarrow\infty}\frac{1}{t}\int^{t}_0\langle\hat O\rangle_{t'}dt'\approx \sum_i|C_i|^2O_{ii} \,.
\label{eq:ergo}
\end{equation}
We use the set of observables $\{\hat M,\hat \theta_{\beta,x}\}$ introduced above to 
sweep the system from an initial state at $F_0<F_{1}$ across the RET regime to a final $F_f>F_{1}$.
Optimal thermalization is then obtained for a sweeping parameter $\lambda\equiv \dot F/d\delta F$ of order 1.
More specifically, we choose a system with ($N=7,L=5$), giving $\mathcal N_s=2288$, and start at $F_0=\Delta/3\pi$ with an
instantaneous eigenstate $|\varepsilon_i(F_0)\rangle$ within the
manifold $M=3$. 
We then compute the microcanonical average $O_{\rm mc}={\Omega^{-1}_{\delta \varepsilon}}\sum\nolimits_i\langle\varepsilon_{i}(F_1)|\hat
O|\varepsilon_{i}(F_1)\rangle$ where $\Omega_{\delta \varepsilon}$ is the number of accessible states within the energy window
$\delta\varepsilon=(0.1-0.2)2\pi F$. The results is shown in Fig.~\ref{fig:3}(a) for our single-particle observable $\hat M$ as well as 
for the two-body correlators $\{\hat \theta_{\beta,x}\}$. All their expectation values converge towards their respective microcanonical
averages via quantum diffusion across the instantaneous spectrum.
For initial states with $M\neq N/2$, these results are confirmed as well, yet the time scale to reach 
thermalization is then typically larger (c.f. also Fig~\ref{fig:2}(c-d)). 

The dependence on $\lambda$ is seen in Fig.~\ref{fig:3}(b). Full delocalization is only reached for $\lambda\sim 1$ in both, energy basis and the TIFB.
For $\lambda\gg 1$, the Shannon entropy, as well as the IPR, strongly depend on the chosen basis; hence the result becomes non-universal and depends on the details of the system.

%******** FIG 4
\begin{figure}[t]
\centering \includegraphics[width=\columnwidth]{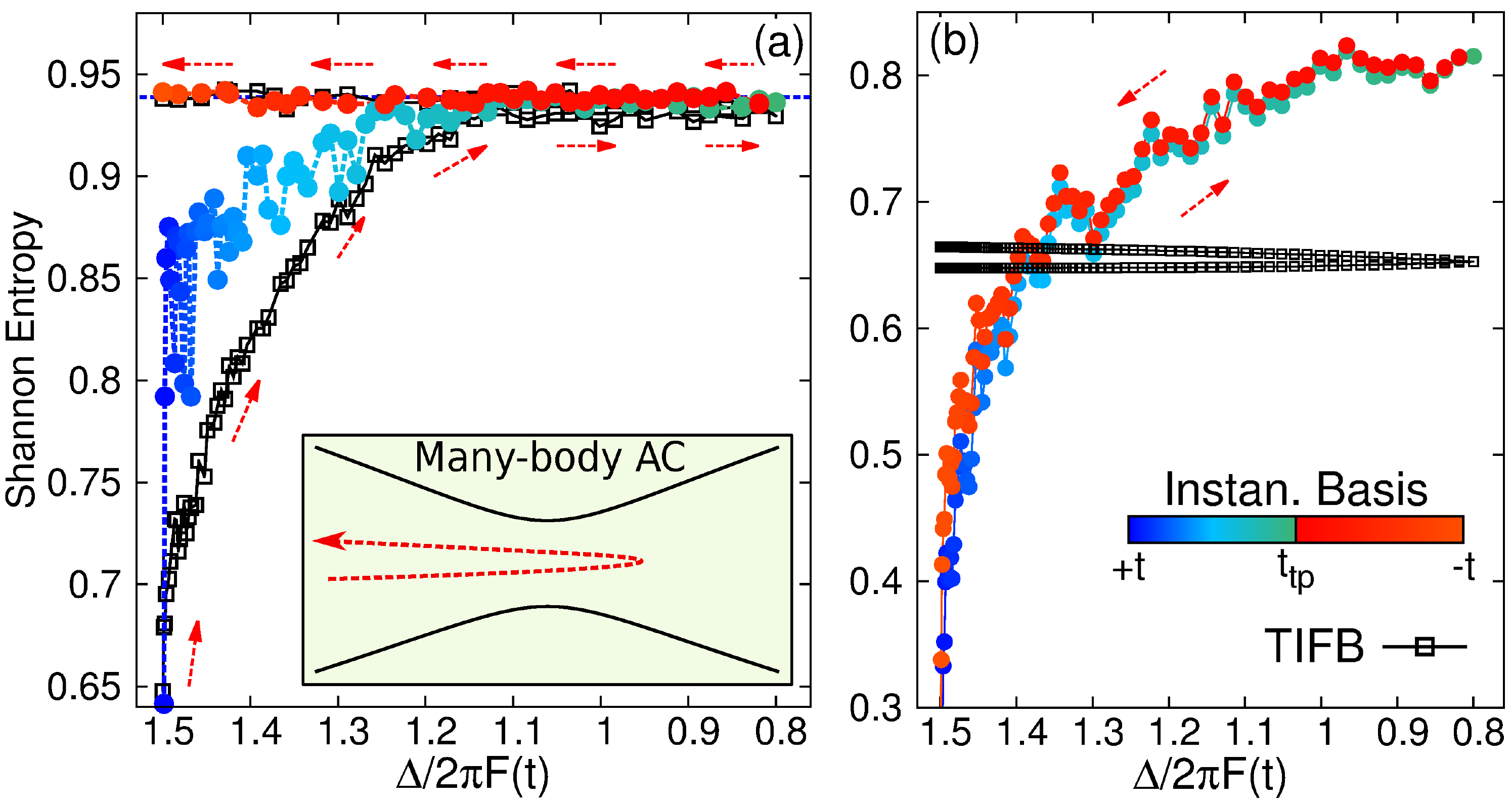}
\caption{\label{fig:4}(Color online): Irreversible (a) and reversible (b) dynamics
via sweeping from $F_0$ to $F_f$  ($F(t)=F_0+\dot F t$) and back ($F(t)=F_f-\dot F t$), characterized by the
Shannon entropy. Entropy in energy basis (color scale: dark-blue/light-red) and  entropy in TIFB (black squares). For (a) $\lambda=1$ and
(b) $\lambda=5\times 10^3$. The initial state is the same as in Fig.\ref{fig:2}(a).}
\end{figure}
%***************
 
Finally, we present an interesting consequence of the just described diffusion process. Since we sweep the force $F(t)$, the system is no
longer autonomous (not even in the Floquet picture) but becomes explicitly (and non-periodically) time dependent. This makes the sweeping
process irreversible in the case of strong (chaotic) thermalization (see Fig.~\ref{fig:4}(a)). However, for fast sweeps with $\lambda \gg 1$,
the process is nearly reversible, see Fig.~\ref{fig:4}(b). Here the system oscillates almost without diffusive spreading between lower and upper band states (e.g., in Fig.~\ref{fig:1}, it would go from the lower left to the upper right states (in black) and back). The latter is similar to the echoes (revivals) in the fidelity as a function of time, yet here in a full many-body context \cite{LoschmidtFidelity}.

\section{Conclusions}

In summary, our two-band system is a paradigm example for the implementation (with ultracold atoms) and the study of complex non-equilibrium quantum evolutions. We showed that one may steer the system into equilibrium or keep it relatively coherent (in the sense of quantum reversibility) depending on the specific choice of quench parameters. This paves the way for future experiments on many-body thermalization \cite{ThermExp1} and new theoretical explorations on optimally controlling the quantum evolution of complex many-body systems \cite{OptcontrolQE,BuchleitScholes2012}.

We acknowledge financial support from the DFG (FOR760) and the HGSFP (GSC 129/1).

\end{document}